\documentclass[manuscript]{aastex}
\usepackage{txfonts}

\shorttitle{Modelling the light curves of PSR B1259-63/LS 2883}
\shortauthors{Kong et al.}

\begin{document}

\title{Modelling the multiwavelength light curves of PSR
B1259-63/LS 2883 - II. The effects of anisotropic pulsar wind and
Doppler-boosting}

\author{S. W. Kong\altaffilmark{1,2}, K. S. Cheng\altaffilmark{2}}
\email{hrspksc@hkucc.hku.hk}

\and

\author{Y. F .Huang\altaffilmark{1,3}}

\altaffiltext{1}{Department of Astronomy, Nanjing University,
Nanjing 210093, China} \altaffiltext{2}{Department of Physics,
University of Hong Kong, Pokfulam Road, Hong Kong}
\altaffiltext{3}{Key Laboratory of Modern Astronomy and Astrophysics
(Nanjing University), Ministry of Education, Nanjing 210093, China}

\begin{abstract}
PSR B1259-63/LS 2883 is a binary system in which a 48-ms pulsar
orbits around a Be star in a high eccentric orbit with a long
orbital period of about 3.4 yr. It is special for having asymmetric
two-peak profiles in both the X-ray and the TeV light curves.
Recently, an unexpected GeV flare was detected by $Fermi$ gamma-ray
observatory several weeks after the last periastron passage. In this
paper, we show that this observed GeV flare could be produced by the
Doppler-boosted synchrotron emission in the bow shock tail. An
anisotropic pulsar wind model, which mainly affects the energy flux
injection to the termination shock in different orbital phase, is
also used in this paper, and we find that the anisotropy in 
the pulsar wind can play a significant role in producing the
asymmetric two-peak profiles in both X-ray and TeV light curves. The
X-ray and TeV photons before periastron are mainly produced by the
shocked electrons around the shock apex and the light curves after
periastron are contributed by the emission from the shock apex and
the shock tail together, which result in the asymmetric two-peak
light curves.
\end{abstract}


\section{Introduction}

The discovery of the PSR B1259-63/LS 2883 system was first reported
in 1992 (Johnston et al. 1992), and it is a binary system containing
a rapidly rotating pulsar, PSR B1259-63, in orbit around a massive
Be star companion LS 2883. The spin period of the pulsar is $P$ =
47.76 ms and the spin-down luminosity is $L_{\rm sd} \simeq 8 \times
10^{35}$ ${\rm ergs}$ ${\rm s}^{-1}$. The distance between the
system and the Earth has been updated to be $2.3 \pm 0.4$ kpc by
Negueruela et al. (2011) recently. The emission from this system has
been widely detected in radio (Johnston et al. 2005), X-rays
(Chernyakova et al. 2006, 2009; Uchiyama et al. 2009) and TeV
gamma-rays (Aharonian et al. 2005, 2009), and the light curves are
modulated on the orbital period. Especially, the X-ray and TeV light
curves are similar and display two-peak profiles. Recently, this
system was first detected in the GeV range by the $Fermi$ satellite
in its last periastron passage in 2010 mid-December (Abdo et al.
2011; Tam et al. 2011). An interesting GeV flare was observed with a
cut-off energy at several hundred MeV, which is difficult to explain
with the traditional lepton model.

In the traditional lepton model of gamma-ray binaries hosting a
pulsar, the interaction between the pulsar wind and the stellar
outflow will terminate the winds with a shock roughly at the
position where the dynamical pressures of the pulsar wind and the
stellar wind are in balance, and this shock can accelerate electrons
to relativistic energies. These accelerated electrons around the
shock apex will emit broadband nonthermal emission via a synchrotron
process for the X-rays or an external inverse Compton (EIC)
scattering of the thermal photons from the Be star for the TeV
gamma-rays (Tavani \& Arons 1997; Dubus 2006; Khangulyan et al.
2007; Takata \& Taam 2009; Kong et al. 2011). The X-ray light curve
reaches a maximum in flux at periastron in the simplest models,
which is inconsistent with observations. Some authors used some
revised leptonic models to explain the drop of photon flux towards
periastron, for example by introducing some non-radiative losses of
electrons (Kangulyan et al. 2007) or varying the microphysical
parameters (Takata \& Taam 2009; Kong et al. 2011). On the other
hand, the synchrotron spectrum has a maximum energy no more than
$2.36 \times 10^8$ eV by equating the synchrotron cooling timescale
with the particle acceleration timescale (see Sect. 2.4), which
seems consistent with the $Fermi$ observed GeV cut-off energy during
the flare. But we should notice that if the acceleration efficiency
is constant, the maximum energy from synchrotron radiation will not
vary along with the orbital phase. If this energy corresponds to the
cut-off energy in the flaring period, the spectra in other periods
cannot be explained properly. The EIC process mainly contributes to
the photons at above 1 GeV, and also cannot explain the observed
flare.

Some previous studies suggested that the GeV flare could be produced
by Doppler-boosting the synchrotron radiation (Kong et al. 2011; Tam
et al. 2011). The interaction between winds in a binary system
should produce a bow-like structure. Bogovalov et al. (2008, 2012)
presented their hydrodynamic simulations of the interaction between
the relativistic and nonrelativistic winds in the PSR B1259-63/LS
2883 system, using both the unmagnetized and magnetized, both the
isotropic and anisotropic pulsar winds. They found that the bulk
motion of the downstream pulsar wind electrons can be accelerated
from a Lorentz factor $\sim 1$ around the shock apex to a very large
Lorentz factor in the bow shock tail. Some previous works
(Khangulyan et al. 2008; Dubus, Cerutti \& Henri 2010) have used the
relativistic Doppler-boosting effect to explain the emission in
gamma-ray binaries. A similar effect should exist in the PSR
B1259-63/LS 2883 system. With a large bulk Lorentz factor, the
emission from the shock tail should be strongly beamed. When the
line-of-sight is near the beaming direction, we can receive the
boosted GeV flux, otherwise, the GeV photons disappear.
Coincidentally, the true anomaly of the GeV flare
($110\textordmasculine - 130\textordmasculine$) is almost the same
as the true anomaly corresponding to the direction of the Earth
($130\textordmasculine$), where the effect of Doppler-boosting is
the most significant. Tam et al. (2011) found that the flux of the
$Fermi$ observations in the flaring period is enhanced by a factor
of 5-10, which suggests a Doppler factor of around 1.5-2. It is
interesting to note that for the X-ray and TeV bands, the second
peaks in the light curves are also around the GeV flaring period.
Therefore these second peaks may be produced by the emission from
the shock apex and the Doppler-boosted emission from the shock tail
together.

The two-peak profiles in both the X-ray and TeV light curves are
also distinctive features of the PSR B1259-63/LS 2883 system. In
addition to the extra-contribution from the Doppler-boosting effect,
some other anisotropic structures in this system may play
significant roles on this problem. Bogovalov \& Khangoulian (2002)
has suggested an anisotropic distribution of energy flux in the
pulsar wind to interpret the torus and jet-like structures in the
center of the Crab Nebula. We can imagine that the anisotropy of
wind is a common phenomenon in pulsars, including PSR B1259-63. If
the spin axis of the pulsar is not perpendicular to the orbital
plane, as the pulsar moving around its companion star, the energy
flux injecting to the termination shock will be modulated with
respect to the orbital phase. This modulation has a two-peak profile
and further produces asymmetric two-peak profiles in the observed
light curves.

In this paper, we will use the Doppler-boosting effect to reproduce
the flare in GeV light curve in detail. We will also use an
anisotropic pulsar wind model, together with the Doppler-boosting
effect, to explain the asymmetric two-peak profiles in both the
X-ray and TeV light curves. A variation of the magnetization
parameter $\sigma$ with the distance to the pulsar suggested in our
previous paper (Kong et al. 2011) is included in our calculations.
The outline of our paper is as follows: in Section 2, we introduce
our model in detail. We then present our results and the comparison
with observations in Section 3. Our discussion and conclusion are
presented in Section 4.

\section{Model Description}

In our model, the broadband emission of the PSR B1259-63/LS 2883
system is mainly from the shock-accelerated electrons, both around
the shock apex and in the shock tail. Due to the interaction between
the pulsar wind and the stellar wind, strong shocks will be formed,
and the electrons (and positrons) can be accelerated at the shock
front of the pulsar wind. This shock will also compress the magnetic
field in the pulsar wind. The shocked relativistic electrons move in
the magnetic field and the photon field of the Be star, and emit
synchrotron and IC radiation to produce the multiband emission.

\subsection{Shock Geometry}

As illustrated in Fig.1, the interaction between the pulsar wind and
the stellar wind will form a shock with a hollow cone-like
structure. The distance from the shock contact discontinuity in the
shock apex to the pulsar can be determined by
\begin{equation}
r_{\rm s}=d \frac{\eta^{1/2}}{1+\eta^{1/2}},
\end{equation}
where $d$ is the separation between the pulsar and its companion and
$\eta$ is the ratio of the momentum fluxes from the pulsar and the
massive star. When the pulsar wind is isotropic, the value of $\eta$
should be $L_{\rm sd} / c \dot{M} v_{\rm w}$, where $c$ is the speed
of light, $\dot{M}$ is the mass-loss rate of the massive star and
$v_{\rm w}$ is the velocity of the stellar wind. The detail of the
wind is presented in Kong et al. (2011). Away from the apex, the
shock surface becomes a hollow cone. The half-opening angle of the
shock contact discontinuity should be (Eichler \& Usov 1993)
\begin{equation}
\theta = 2.1 (1 - \frac{\bar{\eta}^{2/5}}{4}) \bar{\eta}^{1/3},
\end{equation}
where $\bar{\eta} = {\rm min} (\eta,\eta^{-1})$. In this work,
because we use an anisotropic pulsar wind model (see Sec. 2.2), the
value of $\eta$ in the shock apex should vary in different orbital
phases. However, as shown by Bogovalov et al. (2012), the effect of
the anisotropic pulsar wind on the bow-shock structure is relatively
moderate, i.e it can not obviously affect the geometry of the bow
shock, so we use the mean value of $<\eta>$ to determine the
location of the shock apex and the shape of the bow shock in our
calculations. The anisotropy in the pulsar wind mainly affects the
energy flux injection to the termination shock, i.e. affects the
bulk Lorentz factor of the unshocked pulsar wind (see Eq. (3)) and
the downstream magnetic field (see Eq. (7)), in different orbital
phases.

In our model, we approximate that the observed emission is mainly
produced in two regions (as illustrated in Fig.1): (1) Region I
around the shock apex. The bulk motion of the particle flow in this
region is assumed non-relativistic, so the radiation is isotropic.
(2) Region II in the shock tail. As the particle flow propagating
away from the shock apex, the bulk Lorentz factor of the flow will
be increased gradually from $\Gamma_{\rm min} \simeq 1$ in this
region (Bogovalov et al. 2008, 2012). The bulk motion is
mildly-relativistic and the emission should be beamed. When the
line-of-sight is near the beaming direction, as illustrated in Fig.
1, we can receive the Doppler-boosted photons. Note that Eq.(2)
introduced by Eichler \& Usov (1993) is defined at a distance very
far from the shock apex, where the moving directions of the 
unshocked pulsar and stellar winds are nearly parallel. As shown in
Fig. 1, because the moving directions of the  unshocked pulsar
and stellar winds are not the same in Region II, the mean
half-opening angle of the particle flow $\varphi$ should be 
larger than the angle $\theta$ estimated in Eq.(2) and the angle
between the direction of the stellar photons and the beaming
direction should be $ \sim \varphi - \theta$. In this case, even if
the line-of-sight is in the beaming direction, the directions of
photons before and after scattering are not the same, and the EIC
process will not be completely suppressed. When the particle flow is
very far from the shock apex and the massive star, it will deviate
from the flow direction of Region II due to the effects of the
orbital motion and the Coriolis forces (Bosch-Ramon \& Barkov 2011).
So in our work, we assume that the electrons beyond Region II do not
contribute to the observed flux. The bulk Lorentz factor of the flow
at the end of Region II is $\Gamma_{\rm max}$. We also assume that
the electron numbers with different bulk Lorentz factors (from
$\Gamma_{\rm min} \simeq 1$ to $\Gamma_{\rm max}$) are the same in
calculating the synchrotron and EIC radiations.

\subsection{Anisotropic Pulsar Wind}

As suggested by Bogovalov \& Khangoulian (2002), the distribution of
energy flux in the pulsar wind should be anisotropic and the
particle flux can be considered to be more or less isotropic.
Defining the angle between the pulsar spin axis and the direction
from the pulsar to the massive star as $\theta_{\rm PB}$, the
expression of the bulk Lorentz factor of the pulsar wind in the
upstream of the termination shock as a function of $\theta_{\rm PB}$
should be as following:
\begin{equation}
\gamma_1 = \gamma_0 + \gamma_{\rm m} {\rm sin}^2 \theta_{\rm PB},
\end{equation}
where $\gamma_0 \approx 200$, $\gamma_{\rm m} \approx 10^6 - 10^7$.
Then the isotropic particle flux could be described by
\begin{equation}
\dot{N} = \frac{L_{\rm sd}}{m_{\rm e} c^2 (\gamma_0 + \frac{2}{3}
\gamma_{\rm m})},
\end{equation}
where $m_{\rm e}$ is the rest mass of the electron. As the pulsar
orbiting around the massive star, $\theta_{\rm PB}$ will be
modulated on the orbital phase (See Fig. 2), and the characteristics
of the pulsar wind in the termination shock will vary accordingly.

As shown in Khangulyan et al. (2011a, 2011b), the EIC cooling will
reduce the Lorentz factor of the unshocked electrons. We will not
consider this effect in our modelling, and the initial bulk Lorentz
factor parameter of the unshocked pulsar wind $\gamma_{\rm m}$
should be higher if this effect is added. Note that because the
pulsar wind bow shock is extended, we only obtain an upper limit of
the wind anisotropy effect. But the two emission regions in our
model are only a part of the whole bow shock, the effect of the
extending bow shock is relatively moderate. On the other hand, we
can vary the parameter $\theta_{\rm PB}$ to increase the effect of
the anisotropic pulsar wind.

\subsection{Magnetization Parameter}

The magnetization parameter $\sigma$ is defined as the ratio of the
magnetic energy density and the particle kinetic energy density in
the pulsar wind. As estimated in Kong et al. (2011), by using the
pulsar parameters of PSR 1259-63, the magnetization parameter at the
light cylinder should be
\begin{equation}
\sigma_{\rm L} = \frac{B^2_{\rm L}/8\pi}{2\dot{N}_{e^\pm} m_{\rm e}
c/r_{\rm L}^2} \sim 4.68 \times 10^4 (\frac{B_{\rm L}}{2.5 \times
10^4 {\rm G}})^2 (\frac{r_{\rm L}}{2.3 \times 10^8 {\rm cm}})^2
(\frac{N_{\rm m}}{10^4})^{-1} (\frac{\dot{N}_{\rm GJ}}{5.26 \times
10^{31} {\rm s}^{-1}})^{-1},
\end{equation}
where $B_{\rm L}$ is the magnetic field at the light cylinder,
$r_{\rm L}$ is the radius of the light cylinder, $\dot{N}_{e^\pm} =
N_{\rm m} \dot{N}_{\rm GJ}$, $N_{\rm m}$ is the $e^\pm$ multiplicity
and $\dot{N}_{\rm GJ} \sim 5.26 \times 10^{31} (B/3 \times 10^{11}
{\rm G}) (P/47.762 {\rm ms})^{-2} {\rm s}^{-1}$ is the
Goldreich-Julian particle flow at the light cylinder. In outer gap
models (e.g. Cheng, Ho \& Ruderman 1986a, 1986b; Zhang \& Cheng
1997; Takata, Wang \& Cheng 2010), the multiplicity due to various
pair-creation processes could reach $10^4 - 10^5$. Eq. (5) shows
that the pulsar wind is Poynting-dominated at the light cylinder. In
some studies of the Crab Nebula (Kennel \& Coroniti 1984a, 1984b),
the pulsar wind should be kinetic-dominated ($\sigma \sim 0.003$) at
a distance of $r_{\rm s} \sim 3 \times 10^{17}$ cm from the pulsar.
We can imagine that between the light cylinder and the termination
shock, the magnetic energy will be gradually converted into the
particle kinetic energy. In our previous work (Kong et al. 2011), we
have already shown that a variation of magnetization parameter
$\sigma$ with the distance from the pulsar could help us to
reproduce the two-peak profiles in light curves, and we will use the
same variation in this work and describe it as
\begin{equation}
\sigma = \sigma_{\rm L}(\frac{r}{r_{\rm L}})^{-\alpha},
\end{equation}
where $r$ is the distance from the pulsar, and the typical value of
the index $\alpha$ is of the order of unity.

The downstream magnetic field in Region I could be described as
(Kennel \& Coroniti 1984a, 1984b)
\begin{equation}
B=\sqrt{\frac{L_{\rm sd} \sigma}{r^2_{\rm s} c (1 + \sigma)}
\frac{\gamma_1}{\gamma_0 + \frac{2}{3} \gamma_{\rm m}}
(1+\frac{1}{u^2_2})},
\end{equation}
\begin{equation}
u^2_2 =
\frac{8\sigma^2+10\sigma+1}{16(\sigma+1)}+\frac{[64\sigma^2(\sigma+1)^2+20\sigma(\sigma+1)+1]^\frac{1}{2}}{16(\sigma+1)}.
\end{equation}
Because Region II is far from the shock apex and it is an oblique
shock there, the magnetic field should be lower than that in Region
I. We assume the ratio between the magnetic fields in the two
emission regions does not change in different orbital phases, and
the exact ratio is determined by fitting in our work.

\subsection{Radiation Process}

It is usually assumed that the unshocked cold electron pairs in the
pulsar wind can be accelerated to a power-law distribution
$\dot{Q}(\gamma_{\rm e}) \sim (\gamma_{\rm e}-1)^{-p}$ ($\gamma_{\rm
e,min} < \gamma_{\rm e} < \gamma_{\rm e,max}$) in the termination
shock front and be injected into the downstream post-shock flow,
where $p$ is the electron distribution index. The minimum Lorentz
factor can be determined from the conservations of the total
electron number $L_{\rm sd}/\gamma_1 m_{\rm e} c^2=\int
\dot{Q}(\gamma_{\rm e}) {\rm d} \gamma_{\rm e}$ and the total
electron energy $L_{\rm sd}=\int \dot{Q}(\gamma_{\rm e}) \gamma_{\rm
e} m_{\rm e} c^2 {\rm d} \gamma_{\rm e}$, and we can acquire
$\gamma_{\rm e,min}=\gamma_1 (p-2)/(p-1)$ for $p > 2$ (Kirk, Ball \&
Skj{\ae}raasen 1999). The maximum Lorentz factor $\gamma_{\rm
e,max}$ can be determined by equating the cooling timescale of
electrons with the particle acceleration timescale as the following
form,
\begin{equation}
\gamma_{\rm e,max} = \sqrt{\frac{6 \pi e \zeta}{\sigma_{\rm T} B}}
\sim 1.17 \times 10^8 \zeta^{1/2} B^{-1/2},
\end{equation}
where $e$ is the electron charge, $\sigma_{\rm T}$ is the Thompson
scattering cross section, $\zeta$ is the acceleration efficiency
which is usually less than unity. Hereafter the convention $Q_x =
Q/10^x$ is adopted for the cgs units.

The electrons will lose their energies through the radiative or
adiabatic cooling processes, and the evolved electron spectrum
$n(\gamma_{\rm e},t)$ can be obtained from the continuity equation
of the electron distribution (Ginzburg \& Syrovatshii 1964),
\begin{equation}
\frac{\partial n(\gamma_{\rm e},t)}{\partial t}+\frac{\partial
\dot{\gamma}_{\rm e} n(\gamma_{\rm e},t)}{\partial \gamma_{\rm
e}}=\dot{Q}(\gamma_{\rm e}),
\end{equation}
where $\dot{\gamma}_{\rm e}$ is the total energy loss rate of the
electrons and $\dot{Q}(\gamma_{\rm e})$ is the injection rate. The
coefficient of the injection rate in Region I $\eta_{\rm I} =
\dot{Q}(\gamma_{\rm e}) / (\gamma_{\rm e}-1)^{-p}$ ($\gamma_{\rm
e,min} < \gamma_{\rm e} < \gamma_{\rm e,max}$) can be calculated
from $\dot{Q}_{\rm tot} = \int \dot{Q}(\gamma_{\rm e}) {\rm d}
\gamma_{\rm e}$, where $\dot{Q}_{\rm tot} = L_{\rm sd}/[4 m_{\rm e}
c^2 (\gamma_0 + 2 \gamma_{\rm m}/3)]$ by assuming the typical scale
of the shock apex is $r_{\rm s}$ (Dubus 2006). We do not know the
exact structure and physical conditions in Region II, so we assume
the electron injection rates are the same in the two regions for
simplicity. In this case, about half of the pulsar wind electrons
are injected to the emission regions.

Because the cooling and dynamic flow timescales are much smaller
than the orbital period in the PSR B1259-63/LS 2883 system, we use
$\partial n(\gamma_{\rm e},t)/\partial t = 0$ to calculate the
electron distribution at a steady state and acquire (Khangulyan et
al. 2007; Zabalza, Parades \& Bosch-Ramon 2011)
\begin{equation}
n(\gamma_{\rm e}) = \frac{1}{|\dot{\gamma}_{\rm e}|}
\int^{\gamma_{\rm e,max}}_{\gamma_{\rm e}}
\dot{Q}(\gamma^{\prime}_{\rm e}) {\rm d} \gamma^{\prime}_{\rm e}.
\end{equation}
For a source with the dynamical timescale $\tau_{\rm dyn}$, the
electron number at a given Lorentz factor that can accumulate in the
source could be simply calculated by $\dot{Q}(\gamma_{\rm e}) {\rm
min} [\tau_{\rm c}(\gamma_{\rm e}),\tau_{\rm dyn}]$ (Moderski et al.
2005), where $\tau_{\rm c}$ is the cooling timescale. Usually it is
assumed the dynamical timescale in Region I is $\tau_{\rm dyn} = 3
r_{\rm s}/c$. We also assume the dynamical timescale in Region II is
the same with that in Region I for simplicity. For the radiative
cooling timescale $\tau_{\rm c}$, we use the method proposed by
Moderski et al. (2005) to calculate
\begin{equation}
\tau_{\rm c}(\gamma_{\rm e}) = \frac{3 m_{\rm e} c}{4 \sigma_{\rm T}
\gamma_{\rm e} U_B} / [1 + \frac{U_*}{U_B}F_{\rm KN}(\gamma_{\rm
e})],
\end{equation}
where $U_B = B^2/8\pi$ is the magnetic energy density, $U_* = L_{\rm
star} / 4 \pi c R^2$ is the seed photon energy density in Region I
where $L_{\rm star}$ is the luminosity of the massive star and $R$
is the distance between the emission region and the massive star,
$F_{\rm KN} \simeq (1+b)^{-1.5}$ and $b = 4 \gamma_{\rm e} (2.8 k
T_{\rm eff} / m_{\rm e} c^2)$ in Region I, where $k$ is the
Boltzmann constant and $T_{\rm eff}$ is the effective temperature of
the star. In Region II, $U_*$ and $b$ should be reduced by a factor
of $D^2_*$ and $D_*$ respectively, where $D_* = 1 / \Gamma (1-\beta
\cos \theta_*)$ is the Doppler factor, $\Gamma$ is the bulk Lorentz
factor of the particle flow, $\beta = \sqrt{(\Gamma^2-1)}/\Gamma$
and $\theta_*$ is the angle between the direction of the stellar
photons and the moving direction of the flow. Note that a
mono-energetic photon distribution with energy $2.8 k T_{\rm eff}$
is used here as a good approximation of the thermal distribution of
the stellar photons (Moderski et al. 2005).

In our model, the multiband photons from the PSR B1259-63/LS 2883
system are produced by the synchrotron radiation and the EIC process
of the shock-accelerated electrons. An anisotropic inverse-Compton
radiation formula is used in our calculations, in which the
radiation power at frequency $\nu$ from a single electron with
Lorentz factor $\gamma_{\rm e}$ in the comoving frame is given by
(Aharonian \& Atoyan 1981)
\begin{equation}
\frac{{\rm d} P^{\rm EIC}_\nu(\gamma_e,{\rm cos}\theta_{\rm
SC})}{{\rm d} \Omega} = \frac{3 \sigma_{\rm T}}{4 \pi}
\int^\infty_{\nu_{\rm s, min}} {\rm d} \nu_{\rm s} \frac{\nu
f_{\nu_{\rm s}}^{\rm STAR}}{4 \gamma^2_e \nu^2_{\rm s}}
h(\xi,b_\theta),
\end{equation}
\begin{equation}
h(\xi,b_\theta) =
1+\frac{\xi^2}{2(1-\xi)}-\frac{2\xi}{b_\theta(1-\xi)}+\frac{2\xi^2}{b^2_\theta(1-\xi)^2},
\end{equation}
where $h$ is the Planck constant, $\xi=h\nu/(\gamma_{\rm e}m_{\rm
e}c^2)$, $b_\theta=2(1-{\rm cos}\theta_{\rm SC})\gamma_{\rm
e}h\nu_{\rm s}/(m_{\rm e}c^2)$, $h\nu_{\rm s}\ll h\nu \leq
\gamma_{\rm e}m_{\rm e}c^2b_\theta/(1+b_\theta)$, $\theta_{\rm SC}$
is the angle between the injecting photons and the scattered
photons, and is varied along with the orbital phase. In Region II,
this angle in the comoving frame $\theta_{\rm SC}$ can be related to
that in the observer frame $\varphi_{\rm SC}$ by $1- \cos
\theta_{\rm SC} = D_{\rm obs} D_* (1 - \cos \varphi_{\rm SC})$
(Rybicki \& Lightman 1979; Dubus, Gerutti \& Henri 2010; Zdziarski
et al. 2012), where $D_{\rm obs} = 1 / \Gamma (1-\beta \cos
\theta_{\rm obs})$ is the Doppler factor and $\theta_{\rm obs}$ is
the angle between the line-of-sight and the moving direction of the
flow, which is modulated on the orbital period. The flux density of
the massive star photons $f_{\nu_{\rm s}}^{\rm STAR}$ should be $\pi
(R_*/R_{\rm I})^2 2h\nu^3/c^2[{\rm exp}(h\nu/kT_{\rm eff})-1]$ and
$\pi (R_*/R_{\rm II})^2 2D^2_*h\nu^3/c^2[{\rm exp}(D_*h\nu/kT_{\rm
eff})-1]$ (Rybicki \& Lightman 1979; Zdziarski et al. 2012) in
Region I and Region II respectively, where $R_*$ is the radius of
massive star, $R_{\rm I}$ and $R_{\rm II}$ are the distances between
the emission regions and the massive star in Region I and Region II
respectively. Because we do not know the exact structure of Region
II, we use a mean value of $<R_{\rm II}>$ in our calculations. Note
that here the massive star is assumed to be a black body emitter for
simplicity.

The radiation from the electrons in both the shock apex and the
shock tail are included in our calculations. The emission in the
shock apex is assumed isotropic, and the radiation in the shock tail
is beamed. For a point-like source, the Doppler-boosting effect will
increase the photon energy by a factor of $D_{\rm obs}$ and increase
the detected flux by a factor of $D_{\rm obs}^3$ (Dubus, Gerutti \&
Henri 2010). We treat Region II as a ring-like shape on the
cone-like termination shock approximation and we integrate over the
different parts of Region II to get a more accurate result.
According to Eq.(9), the synchrotron spectrum will cut off at the
energy of
\begin{equation}
h \nu_{\rm syn,max} (\theta_{\rm obs}) = D_{\rm obs} (\theta_{\rm
obs}) \frac{3 h e \gamma_{\rm e,max}^2 B}{4 \pi m_{\rm e} c} \sim
2.36 \times 10^8 \zeta D_{\rm obs} (\theta_{\rm obs}) {\rm eV}.
\end{equation}
The variation of $D_{\rm obs}$ will help us to obtain the observed
cut-off energy in the flaring period and the flare structure in the
light curve.

\section{Results}

In this section, we will present some calculated results using our
model, and compare them with the observations. We use the updated
parameters for the PSR B1259-63/LS 2883 system in our calculations
as follows (Negueruela et al. 2011): For the orbital parameters, we
take the eccentricity $e = 0.87$, the semimajor axis $a = 7$ AU; For
the compact object PSR B1259-63, we take the spin-down luminosity
$L_{\rm sd} = 8 \times 10^{35}$ $\rm erg$ $\rm s^{-1}$; For the Be
star LS 2883, we take the stellar luminosity $L_{\rm star} = 7.3
\times 10^4 L_\odot$, the stellar radius $R_* = 10 R_\odot$, the
effective temperature of the star $T_{\rm eff} = 30000$ $\rm K$. The
distance between the system and the Earth is taken to be 2.3 kpc.
The angle between the line-of-sight and the orbital plane is taken
to be $65\textordmasculine$, and the true anomaly corresponding to
the direction of the Earth is taken to be $130\textordmasculine$.
The input parameters we used are as follows: The mean ratio of the
momentum fluxes from the pulsar and the massive star is $<\eta> =
0.16$, which corresponds to the mass-loss rate of the massive star
$\dot{M} \sim 2.6 \times 10^{-8}$ $M_{\sun}$ ${\rm yr}^{-1}$ with a
velocity of $v_{\rm w} \sim 10^8$ ${\rm cm}$ ${\rm s}^{-1}$, which
is consistent with the typical value of the polar wind in the Be
star (Waters et al. 1988). The corresponding half-opening angle of
the shock cone is $ \theta \sim 58\textordmasculine$. We choose the
half-opening angle of Region II $\varphi = 65\textordmasculine$, so
that around the true anomaly $130\textordmasculine$, we can receive
strong beamed emission from the shock tail. The angle between the
direction of stellar photons and the moving direction of the flow in
Region II is taken as $\theta_* \sim \varphi - \theta$. The maximum
bulk Lorentz factor contributing to the observed flux in the shock
tail is $\Gamma_{\rm max} = 2.0$, the bulk Lorentz factor parameters
of the unshocked pulsar wind are $\gamma_0 = 200$ and $\gamma_{\rm
m} = 2.0 \times 10^6$, the true anomaly of the projection of
the pulsar spin axis in the orbital plane is $20\textordmasculine$,
the angle between the pulsar spin axis and the orbital plane is
$46\textordmasculine$, the magnetization parameter at the light
cylinder is $\sigma_{\rm L} = 8 \times 10^3$ (for $N_{\rm m} \sim
5.8 \times 10^4$) and the decay index is $\alpha = 1.1$, the
electron distribution index is $p = 2.1$, the acceleration
efficiency is $\zeta = 0.36$. The ratio of magnetic fields between
Region II and Region I is assumed to be 0.1. The mean distance
between the massive star and Region II $<R_{\rm II}>$ is taken to be
2.0 times of the binary separation, which is consistent to the
results in Bogovalov et al. (2008, 2012) for $\Gamma_{\rm max} =
2.0$. Note that by using the parameters introduced above, the
dynamical timescale $\tau_{\rm dyn}$ in Region I is $\sim d/c$ and
that in Region II is less than $\sim 2 d/c$. The difference between
these two timescales is small, so the assumption that these two
regions have the same dynamical timescales in our calculations is
reasonable. The periastron is taken to be at orbital phase $\sim 0$
throughout the paper.

In Fig. 2, we show the variations of the angle between the pulsar
spin axis and the line joining the two stars $\theta_{\rm BP}$, the
bulk Lorentz factor of the unshocked pulsar wind $\gamma_1$, the
magnetization parameter at the termination shock $\sigma$, and the
magnetic field around the shock apex $B$ with respect to the orbital
phase. We can see that  because the pulsar spin axis is not
perpendicular to the orbital plane, $\theta_{\rm BP}$ is modulated
on the orbital phase. For the case that the true anomaly of the
projection of the pulsar spin axis in the orbital plane is
$20\textordmasculine$ and the angle between the pulsar spin axis and
the orbital plane is $46\textordmasculine$, $\theta_{\rm BP}$ will
reach the minimum and maximum of $46\textordmasculine$ and
$134\textordmasculine$ at the true anomaly of $20\textordmasculine$
and $-160\textordmasculine$ respectively. By using the anisotropic
energy flux injection of the pulsar wind described in Section 2.2,
 $\gamma_1$ will vary with respect to $\theta_{\rm BP}$ (See Eq.
(3)) and reach the maximum when $\theta_{\rm BP} =
90\textordmasculine$ at the true anomaly of $-70\textordmasculine$
and $110\textordmasculine$. In this case a two-peak profile appears
in the distribution of $\gamma_1$. This two-peak structure will
affect the minimum Lorentz factor $\gamma_{\rm e,min}$ in the
electron distribution, and further affects the electron number at a
certain Lorentz factor. The two-peak distribution on the electron
number at a certain Lorentz factor will help us to obtain the
two-peak profiles in light curves. We can also see that the
magnetization parameter at the termination shock $\sigma$ is
modulated with respect to orbital phase, which is due to the
different shock distance $r_{\rm s}$ in different orbital phases.
The magnetic field in the shock apex $B$ is also modulated with
respect to orbital phase, which is similar to the modulation of
magnetization parameter $\sigma$. But this modulation is not
symmetrical because of the anisotropy of the pulsar wind (See Eq.
(3) and Eq. (7)).

Here we use the parameters introduced above to calculate the spectra
and the comparisons with observations are shown in Fig. 3. The
related timescales are shown in Fig. 4. It can be seen that our
calculated X-ray and GeV emission is mainly produced by the
synchrotron radiation and the TeV photons are mainly contributed by
the EIC effect where the relativistic shocked electrons up-scatter
the soft photons from the massive star. The X-ray and TeV flux at
true anomaly of $120\textordmasculine$ is contributed by the
emission from shock apex and shock tail together, but the flux at
true anomaly of $-60\textordmasculine$ is mainly produced by the
electrons in the shock apex. By choosing the acceleration parameter
$\zeta = 0.36$ in our modelling, the cut-off energy of the
synchrotron spectrum in the unboosted region is $h \nu_{\rm syn,max}
\sim 84$ MeV, which is a little lower than the lower limit of the
energy range of the observations (100 MeV). So in the lower panel of
Fig. 3, the pre-periastron synchrotron spectrum only have a small
contribution in the above 100 MeV range. In the upper panel of Fig.
3, the line-of-sight is near the beaming direction. The photon flux
and the photon energy will be strongly Doppler-boosted, and the
cut-off energy in the synchrotron spectrum could be boosted to $h
\nu_{\rm syn,max} \sim 300$ MeV, consistent with the $Fermi$
observed GeV cut-off energy during the flare (Abdo et al. 2011; Tam
et al. 2011). In this case, the observed flare can be reproduced.
Unlike in the synchrotron component above 100 MeV, the EIC component
above 1 TeV is not dominated by the boosted emission from the shock
tail in the upper panel of Fig. 3. This is because (1) in the
comoving frame the seed photon density in the shock tail is lower
than that in the shock apex reduced by the Doppler de-boosting; (2)
because of the anisotropy of the EIC emission, the flux will be
suppressed when the angle between the directions of the input and
output photons is small. In Fig. 4, we can see that the radiative
cooling of the electrons with Lorentz factor $10^5-10^6$ is mainly
dominated by the EIC radiation in the Klein-Nishina (KN) regime, and
the radiative cooling of the electrons with higher energies is
dominated by the synchrotron radiation. In our calculations, the
minimum Lorentz factor of the shocked electrons is $\sim 10^5
\gamma_{1,6}$, so the EIC process will never be in the Thomson
regime. As a result, the synchrotron component is higher than the
EIC component in our calculated spectra. Our calculated synchrotron
spectrum in the lower panel of Fig.3 does not fit the observations
in the GeV range well, but note that the spectrum data in
pre-periastron period given by different groups (Abdo et al. 2011;
Tam et al. 2011) also are not consistent with each other, which may
be due to the low value of the photon flux.

Our calculated broadband light curves of the PSR B1259-63/LS 2883
system and the comparisons with observations are presented in Fig.
5. We can see that the asymmetric two-peak profiles in the X-ray and
TeV light curves can be well reproduced by our model. The
observations before periastron are mainly contributed by the
emission in the shock apex. The shock apex region also contributes
to some emission in the post-periastron range, and the whole light
curves display the asymmetric two-peak profiles with the flux before
periastron higher than that after periastron. The anisotropic pulsar
wind plays a significant role in producing this asymmetric two-peak
profile because of its anisotropic energy flux injection to the
termination shock. The emission from the shock tail is unimportant
before periastron, because the angle between the line-of-sight and
the direction of the flow in the shock tail $\theta_{\rm obs}$ is
large. But it has an obvious contribution in the light curves after
periastron, i.e. the second peaks in both the X-ray and TeV light
curves are produced by the electrons in the shock apex and the shock
tail together. For the GeV light curves, we can see that almost all
the flux in the GeV flares are produced by the emission from the
shock tail, where the emission is beamed and both the photon energy
and detected flux are Doppler-boosted.

\section{Conclusions and Discussion}

The PSR B1259-63/LS 2883 system is an attractive binary system and
special for having distinct modulations in the light curves of
different energy bands. The X-ray and TeV light curves are similar
and have two peaks before and after periastron respectively. The GeV
light curve has an interesting flare several weeks after the
periastron passage, whose spectrum cuts off at several hundred MeV.
In this paper, we have modelled the X-ray, GeV and TeV observations
of this system. In our calculations, an anisotropic pulsar wind
model and the relativistic Doppler-boosting effect are considered,
and the X-ray and GeV photons are mainly from the synchrotron
radiation and the TeV emission is mainly from the EIC process. We
also assume that the emission from both the shock apex and the shock
tail can contribute to the observations, and the shock apex
radiation is isotropic and the emission from the shock tail is
beamed because of the flow in the tail moving
mildly-relativistically. We find that for the X-rays and TeV
gamma-rays, the anisotropic energy flux injection of the pulsar wind
plays an important role in producing the asymmetric two-peak
profiles in the light curves. The observations before the periastron
are mainly produced by the shocked electrons around the shock apex
and the photons after periastron are contributed by the emission
from the shock apex and the shock tail together. For the GeV
gamma-rays, the observed flare is contributed by the emission from
the shock tail, where the synchrotron photons are Doppler-boosted
strongly. Unfortunately, we do not know the exact structure and
physical processes in the shock tail, so in our calculations we make
some assumptions, and we just show that our model is a possible way
to reproduce the multiwavelength features in the PSR B1259-63/LS
2883 system. A more exact modelling could be done by using the
detailed simulation results on binary pulsar systems (Bogovalov et
al. 2008, 2012; Takata et al. 2012).

In our modelling, we use an assumption that the magnetization
parameter $\sigma$ varies with the distance to the pulsar, which is
in principle possible. Some recent studies found that the variation
of magnetization parameter $\sigma$ could be in a different way. For
example, Aharonian, Bogovalov \& Khangulyan (2012) showed that the
magnetization parameter $\sigma$ should decrease abruptly within
$10^{10}$ cm from the pulsar by fitting the observations of the Crab
pulsar. Because the exact conversion process from Poynting flux to
kinetic energy in the pulsar wind is still unclear, we think both
models cannot be excluded. Some previous models implied that the
energy conversion could exist over the entire distance from the
pulsar to the termination shock (Coroniti 1990; Lyubarsky \& Kirk
2001), and Contopoulos \& Kazanas (2002) further showed that
$\sigma$ could decrease inversely proportional to the distance from
the light cylinder. In our previous paper (Kong et al. 2011), a
variation of the magnetization parameter $\sigma$ with the distance
from the pulsar could help us to reproduce the two-peak profiles in
light curves, and we also need a variation index $\alpha = 1.1$ to
fit the observations better in this work. The adoption of $\alpha =
1.1$ here is purely phenomenological, which is affected by both the
unclear magnetic energy dissipation and the EIC cooling of the
unshocked pulsar wind (Khangulyan et al. 2011a, 2011b).

In our previous paper (Kong et al. 2011), we discussed the effect of
the disk in the stellar wind on the X-ray and TeV light curves. The
existence of a disk in the PSR B1259-63/LS 2883 system was confirmed
by the radio observations (Johnston et al. 1996, 2005), but the
exact position of the disk is still unclear. Some radio observations
suggested that the disk is tilted with respect to the orbital plane
and the line of intersection between the disk plane and the orbital
plane is oriented at about $90\textordmasculine$ with respect to the
major axis of the binary orbit (Wex et al. 1998; Wang, Johnston \&
Manchester 2004). Chernyakova et al. (2006) further suggested that
the half-opening angle of the disk (projected on the pulsar orbital
plane) is $\Delta \theta_{\rm disk} \simeq 18\textordmasculine.5$,
and the intersection between the stellar equatorial plane and the
orbital plane is inclined at $\theta_{\rm disk} \simeq
70\textordmasculine$ to the major axis of the pulsar orbit by
fitting the X-ray and TeV light curves. Our previous calculations
(Kong et al. 2011) showed that the X-ray flux increases in the
passage of the disk, but the flux in the TeV range decreases
significantly, which is consistent with the analysis by Kerschhaggl
(2011). But note that we only consider the emission from the shock
apex in Kong et al. (2011). When the pulsar entering the disk, the
mass flux density will increase by a factor of 30-100 (Waters et al.
1988). By assuming the stellar wind velocity reduced by a factor of
10, the momentum flux density ratio $\eta$ will decrease by a factor
of 3-10 in the disk. In this case, the half-opening angle of the
emission region in the shock tail will decrease and the angle
between the line-of-sight and the direction of the flow in the shock
tail $\theta_{\rm obs}$ will increase. The effect of
Doppler-boosting will be suppressed and the emission from shock tail
will be reduced. We can see from Fig. 5 that there are only upper
limits in GeV light curves between true anomaly
$90\textordmasculine$ and $115\textordmasculine$. The non-detection
in this period may be due to the suppression of the shock tail
emission in the disk. The TeV observations also show a reduction in
flux during the estimated disk passage, which are consistent with
the explanation that emission from both the shock apex (Kong et al.
2011) and the shock tail will decrease in the disk. In X-ray band,
there are significant signals between true anomaly
$90\textordmasculine$ and $115\textordmasculine$. But note that the
X-ray flux from the shock apex could increase in the passage of the
disk (Kong et al. 2011). The increase of the shock apex emission and
the decrease of the shock tail emission in the disk will compete
with each other, and the total flux could have no significant
change.

In a recent paper, Khangulyan et al. (2011a) investigated the
emission spectrum produced by the EIC process of the unshocked
pulsar wind. They argued that the gamma-ray flare in GeV band after
the periastron can be explained by the EIC emission of the cold
pulsar wind with the bulk Lorentz factor $\gamma \approx 10^4$. The
Be star disk plays a significant role in producing GeV flare in
their model. First, the radiation of the shocked stellar disk can
provide a dense photon target for the EIC scattering (van Soelen \&
Meintjes 2011). Second, the strong ram pressure inside the disk
makes the wind termination shock stand close to the pulsar, and then
the EIC luminosity should be suppressed. When the pulsar escapes the
disk, the unshocked pulsar wind zone towards the observer is
significantly increased. Consequently, an enhancement of the
gamma-rays will be observed. Although in our model the GeV flare is
produced by the Doppler-boosted synchrotron radiation, the EIC
process of the unshocked pulsar wind may also have some
contributions. Note that in the pre-periastron spectrum data in Tam
et al. (2011), the $Fermi$ detected emission is concentrated in a
narrow band between 1-25 GeV. The observed spectrum is different
from the spectrum produced by synchrotron radiation, and is similar
to the calculated spectra in Khangulyan et al. (2011a). If the data
in this range is indeed correct, we argue that this part of GeV
photons may be produced by the EIC process of the unshocked
electrons before termination shock in the pulsar wind.

If the 3D structure of the shock is a simple hollow cone at the
shock tail, we expect that two flares per orbital period will be
observed when the angle between the line-of-sight and the orbital
plane is smaller than the half-opening angle of the shock cone, and
each flare can last for about $2/\Gamma$ orbital phase $\sim
12^{\circ}(\Gamma/10)^{-1}$, where $\Gamma$ is the bulk Lorentz
factor at the tail of the shock. Otherwise, when the line-of-sight
is near the edge of the hollow cone or outside the hollow cone, as
illustrated in Fig. 1, only one flare could be observed in the light
curve. As the angle between the line-of-sight and the flow direction
increasing, the flare will be smoother and disappear eventually. In
principle, our model could be used in other similar gamma-ray binary
systems in our Galaxy with different observational angles and
distinct light curves.

\acknowledgments

We would like to thank the anonymous referee for stimulating
suggestions that lead to an overall improvement of this study. We
also would like to thank Y. W. Yu for helpful suggestions and
discussion. This research was supported by a 2011 GRF grant of the
Hong Kong Government under HKU700911p. YFH was supported by the
National Natural Science Foundation of China (Grant No. 11033002)
and the National Basic Research Program of China (973 Program, Grant
No. 2009CB824800).

\clearpage

\begin{figure}
\epsscale{1.0} \plotone{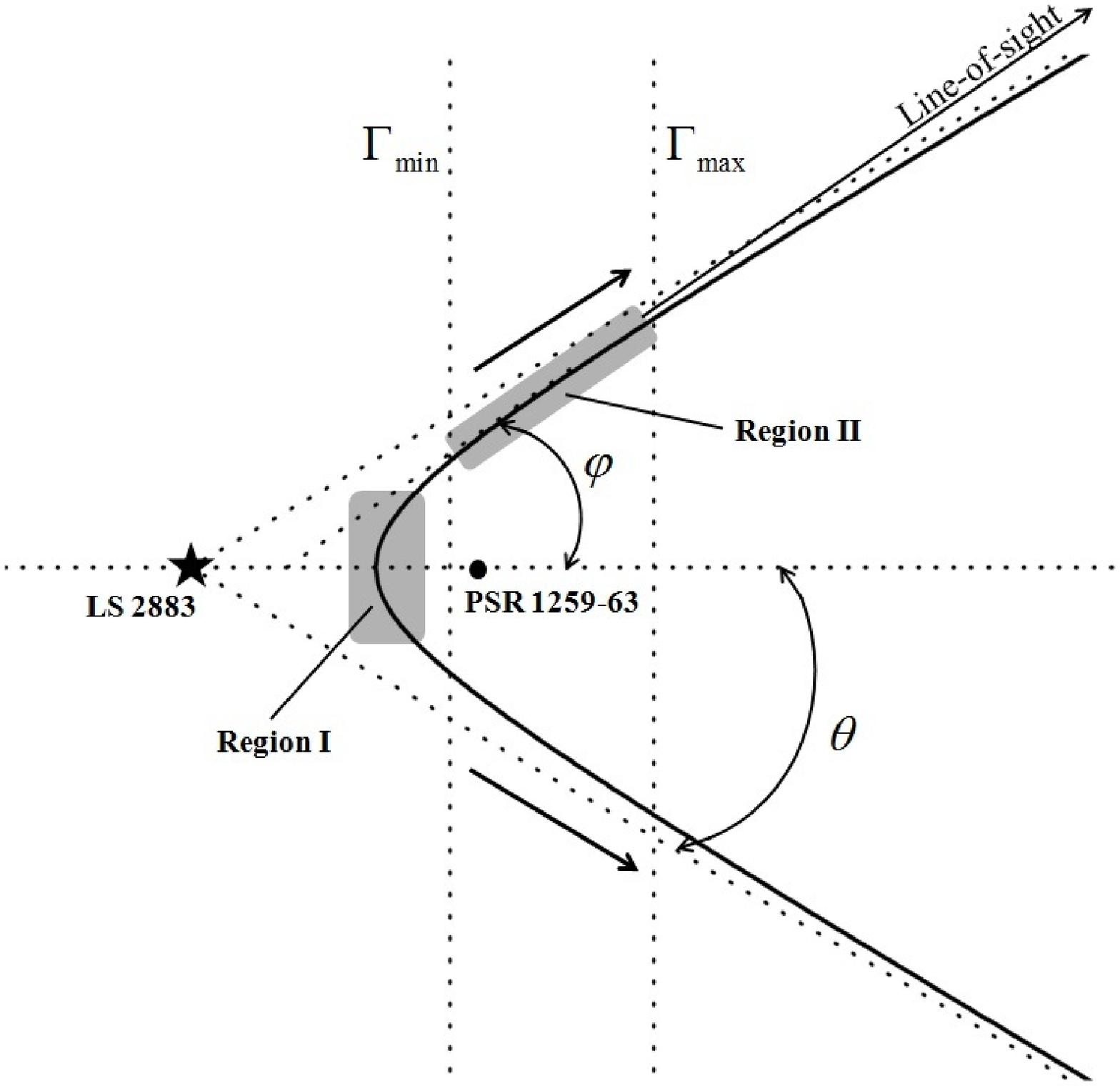} \caption{Geometry of the termination
shock. The interaction between the pulsar wind and the stellar wind
forms a termination shock with a hollow cone-like structure. Region
I is around the shock apex, and the particle flow there is moving
non-relativistically and radiating isotropically. Region II is in
the shock tail, and the particle flow there is moving
mildly-relativistically and the emission is beamed.\label{fig1}}
\end{figure}

\clearpage

\begin{figure}
\epsscale{0.7} \plotone{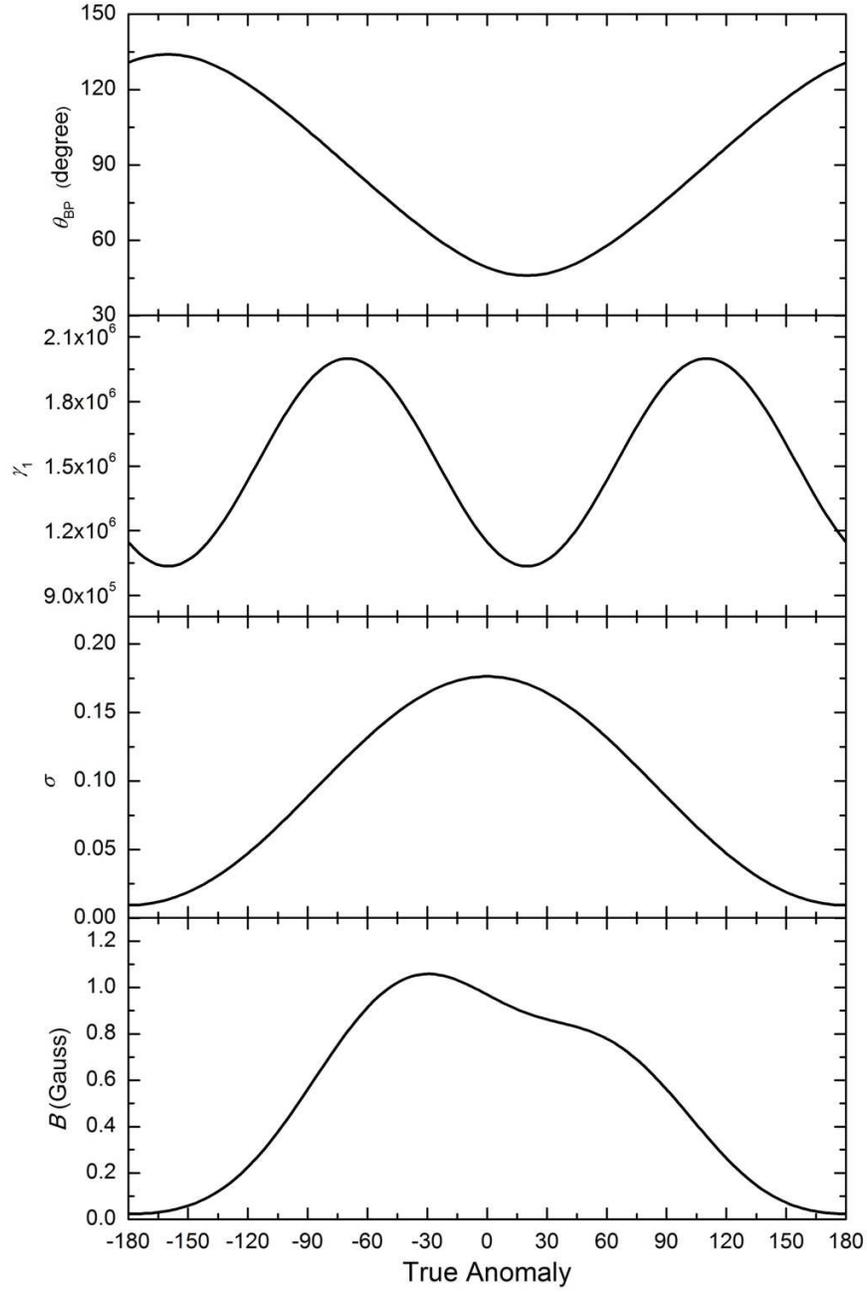} \caption{From top to bottom: the
variations of the angle between the pulsar spin axis and the line
joining the two stars $\theta_{\rm BP}$, the bulk Lorentz factor of
the unshocked pulsar wind $\gamma_1$, the magnetization parameter at
the termination shock $\sigma$ and the magnetic field around the
shock apex $B$ with respect to the orbital phase. \label{fig2}}
\end{figure}

\clearpage

\begin{figure}
\epsscale{0.8} \plotone{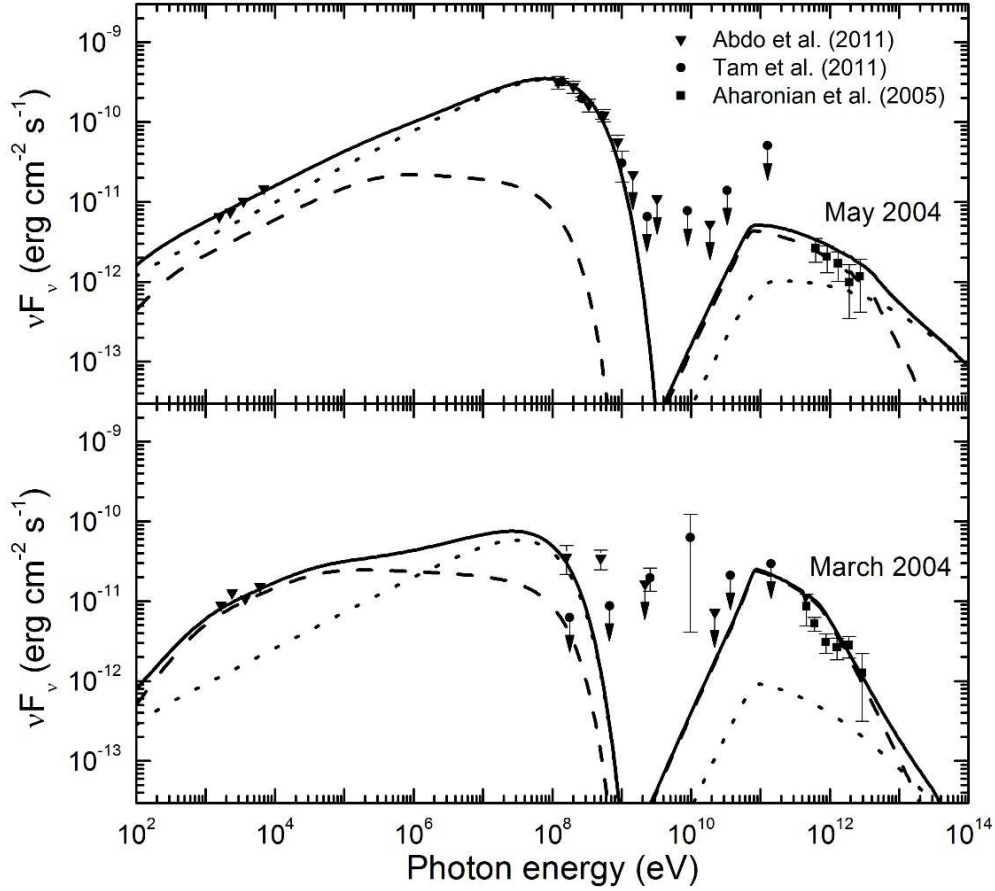} \caption{The calculated spectra as
compared with observations. The spectrum data are taken from
Aharonian et al. (2005), Abdo et al. (2011) and Tam et al. (2011).
In the upper panel, the spectra are calculated at true anomaly of
$120\textordmasculine$ and the data are taken in the post-periastron
range. In the lower panel, the spectra are calculated at true
anomaly of $-60\textordmasculine$ and the data are taken in the
pre-periastron range. The dashed lines and dotted lines correspond
to the emission from the shock apex and shock tail respectively, and
the solid lines correspond to the total flux. \label{fig3}}
\end{figure}

\clearpage

\begin{figure}
\epsscale{0.8} \plotone{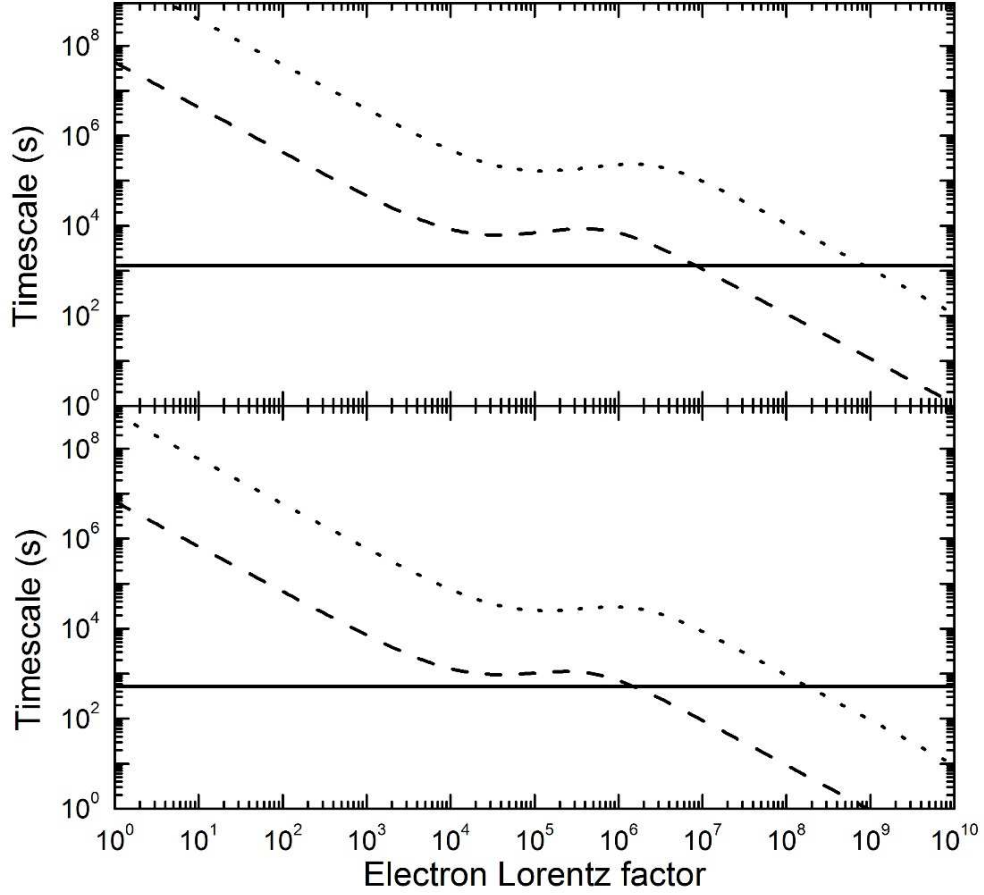} \caption{The calculated timescales
with respect to the electron energies. The solid lines correspond to
the dynamical timescale $\tau_{\rm dyn}$, the dashed lines and
dotted lines correspond to the cooling timescales $\tau_{\rm c}$ in
the shock apex and shock tail respectively. In the upper panel, the
timescales are calculated at true anomaly of $120\textordmasculine$,
and in the lower panel, the timescales are calculated at true
anomaly of $-60\textordmasculine$. \label{fig4}}
\end{figure}

\clearpage

\begin{figure}
\epsscale{0.7} \plotone{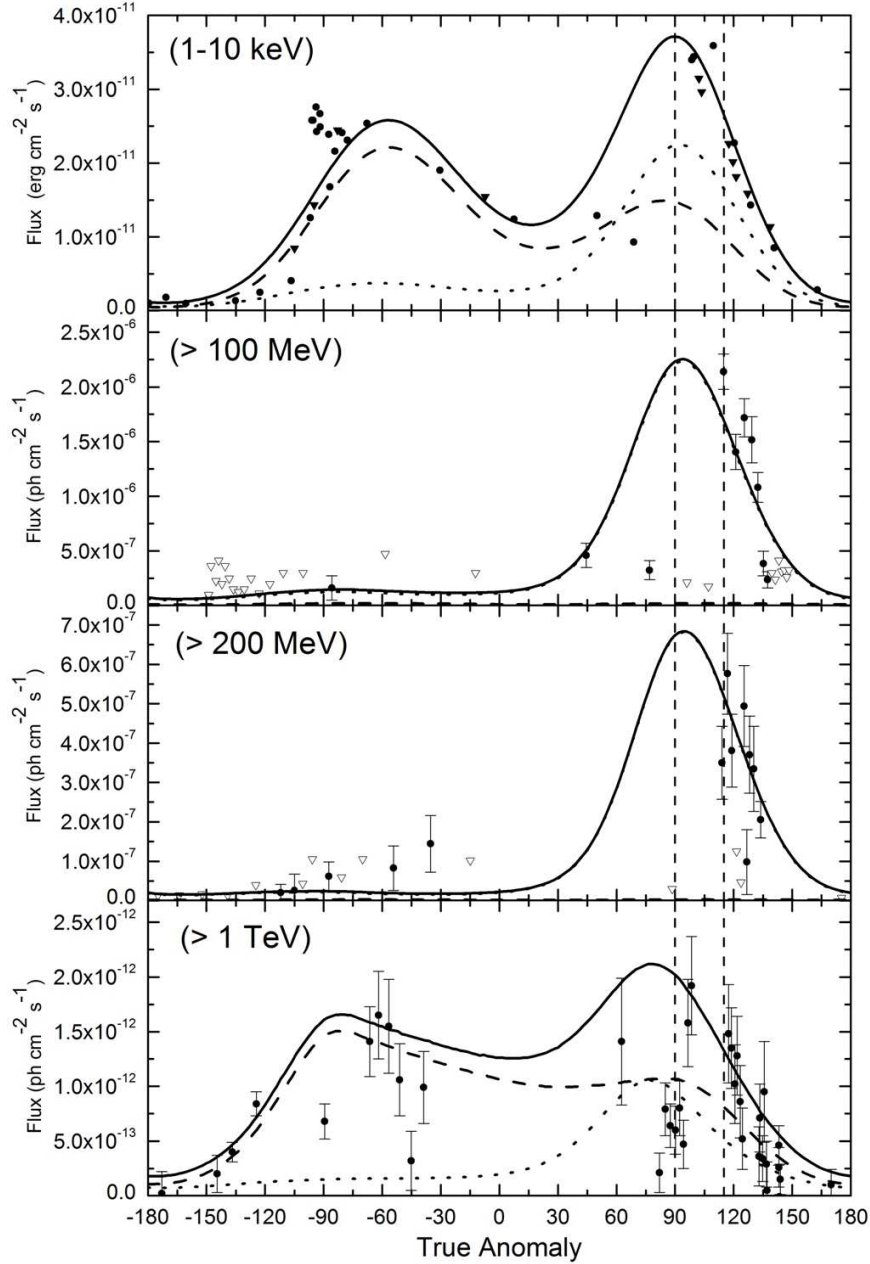} \caption{The calculated multi-band
light curves as compared with observations. The X-ray light curve
data are taken from Chernyakova et al. (2006, 2009; circles) and
Abdo et al. (2011; triangles), the $>$ 100 MeV light curve data are
taken from Abdo et al. (2011), the $>$ 200 MeV light curve data are
taken from Tam et al. (2011) and the TeV light curve data are taken
from Aharonian et al. (2005, 2009). The dashed lines and dotted
lines correspond to the emission from the shock apex and shock tail
respectively, and the solid lines correspond to the total flux. The
empty triangles are upper limits in observations.  The vertical
dashed lines correspond to the estimated disk passage. \label{fig5}}
\end{figure}

\clearpage

\end{document}